\documentstyle[prl,floats,aps,twocolumn,epsf,graphicx]{revtex}
\begin{document}
\twocolumn[\hsize\textwidth\columnwidth\hsize\csname
@twocolumnfalse\endcsname

\title{Quasinormal Spectrum and Quantization of Charged Black Holes}
\author{Shahar Hod}
\address{The Racah Institute of Physics, The Hebrew University, Jerusalem 91904, Israel}
\address{and}
\address{The Ruppin Academic Center, Emeq Hefer 40250, Israel}
\date{\today}
\maketitle

\begin{abstract}

\ \ \ Black-hole quasinormal modes have been the subject of much recent attention, 
with the hope that these oscillation frequencies may shed some light on 
the elusive theory of quantum gravity. We study {\it analytically} the asymptotic quasinormal 
spectrum of a {\it charged} scalar field in the (charged) Reissner-Nordstr\"om spacetime. 
We find an analytic expression for these black-hole resonances in terms of the black-hole 
physical parameters: its Bekenstein-Hawking temperature $T_{BH}$, 
and its electric potential $\Phi$. 
We discuss the applicability of the results in the context of black-hole quantization. 
In particular, we show that according to Bohr's correspondence principle, 
the asymptotic resonance corresponds to a fundamental area unit $\Delta A=4\hbar\ln2$.
\end{abstract}
\bigskip

]

Everything in our past experience in physics tell us that general relativity and quantum theory are 
approximations, special limits of a single, universal theory. 
However, despite the flurry of research in this field we still lack a complete theory of quantum gravity.
In many respects the black hole plays the same role in gravitation that the atom 
played in the nascent of quantum mechanics \cite{Bekenmar}. 
It is therefore believed that black holes may play a major role in our 
attempts to shed light on the nature of a quantum theory of gravity.

The quantization of black holes was proposed long ago by Bekenstein
\cite{Beken1,Beken2}, based on the remarkable observation that the horizon
area of a non-extremal black hole behaves as a classical 
adiabatic invariant. In the spirit of the Ehrenfest principle
\cite{Ehren} -- any classical adiabatic invariant
corresponds to a quantum entity with a discrete spectrum, and based on the idea of a minimal increase in 
black-hole surface area \cite{Beken1}, Bekenstein conjectured that the horizon area of a quantum
black hole should have a discrete spectrum of the form

\begin{equation}\label{Eq1}
A_n=\gamma {\ell^2_P} \cdot n\ \ \ ;\ \ \ n=1,2,3,\ldots\ \  ,
\end{equation}
where $\gamma$ is a dimensionless constant, and 
$\ell_P=(G\hbar/c^3)^{1/2}$ is the
Planck length (we use units in which $G=c=\hbar=1$ henceforth). 
This type of area quantization has since been reproduced based on various other considerations 
(see e.g., \cite{Hod1} for a detailed list of references). 

In order to determine the value of the coefficient $\gamma$, 
Mukhanov and Bekenstein \cite{Muk,BekMuk,Beken3} have suggested, 
in the spirit of the Boltzmann-Einstein formula in statistical physics, to 
relate $g_n \equiv \exp[S_{BH}(n)]$ to the number of the black hole microstates 
that correspond to a particular external macro-state, where $S_{BH}$ is the black-hole entropy. 
In other words, $g_n$ is the degeneracy of the $n$th area eigenvalue. 
Now, the thermodynamic relation between black-hole surface area and entropy, $S_{BH}=A/4\hbar$, 
can be met with the requirement that $g_n$ has to be an integer for every $n$ only when
 
\begin{equation}\label{Eq2}
\gamma =4\ln{k} \  ,
\end{equation}
where $k$ is some natural number. 

Identifying the specific value of $k$ requires further input. This information 
may emerge by applying {\it Bohr's correspondence principle} to 
the (discrete) quasinormal mode (QNM) spectrum of black holes \cite{Hod2}. 
Gravitational waves emitted by a perturbed black hole are dominated by this 
`quasinormal ringing', damped oscillations with a {\it discrete}
spectrum (see e.g., \cite{Nollert1} for a detailed review). 
At late times, all perturbations are
radiated away in a manner reminiscent of the last pure dying tones of
a ringing bell \cite{Press,Cruz,Vish,Davis}. 
These black-hole resonances are the characteristic `sound' of
the black hole itself, depending on its parameters: mass, charge and angular momentum.

It turns out that for a Schwarzschild black hole, for a given angular harmonic index $l$ there exist 
an infinite number of (complex) quasinormal frequencies, characterizing oscillations with decreasing 
relaxation times (increasing imaginary part) \cite{Leaver,Bach}. 
On the other hand, it was found numerically  \cite{Leaver,Nollert2,Andersson} that 
the real part of the Schwarzschild gravitational resonances approaches an 
asymptotic constant value. 
Based on Bohr's correspondence principle, it was argued \cite{Hod2} that 
the asymptotic resonances are given by \cite{Note1} 
(we assume a time dependence of the form $e^{-i\omega t}$),

\begin{equation}\label{Eq3}
\omega = \pm T^{s}_{BH}\ln3 -i 2\pi T^{s}_{BH} (n+{1\over2})\  ,
\end{equation}
where $T^{s}_{BH}=1/8\pi M$ is the Bekenstein-Hawking temperature of the Schwarzschild black hole. 
An analytical proof of this equality was later given in \cite{Motl}. 

The emission of a quantum of frequency $\omega$ results in a change $\Delta M=\hbar \omega_R$ in 
the black-hole mass. Assuming that $\omega$ corresponds to the asymptotically damped 
limit Eq. (\ref{Eq3}) \cite{Note1}, and using the first-law of black-hole thermodynamics 
$\Delta M={1 \over 4}T^{s}_{BH} \Delta A$, this implies 
a change $\Delta A=4\hbar \ln3$ in the black hole surface area. 
Thus, the correspondence principle, as applied to the black-hole resonances, 
provides the missing link, and gives evidence in favor of the value $k=3$. 

Furthermore, it was later suggested to use the black-hole 
QNM frequencies in order to fix the value of the Immirzi parameter in Loop Quantum Gravity, 
a viable approach to the quantization of general relativity \cite{Dreyer,Ash,Rov}. 
The intriguing proposals outlined above \cite{Hod2,Dreyer} have triggered a flurry of research 
attempting to calculate the asymptotic ringing frequencies of various types of 
black holes (for a detailed list of references see, e.g., \cite{HodKesh}).

It should be emphasized however, that former analytical studies of the asymptotic QNM spectrum 
did not include {\it chemical potentials}, such as rotation or electric charge. 
These potentials enter into the first law of black-hole thermodynamics for 
rotating black holes, or for a charged scalar field in the (charged) 
Reissner-Nordstr\"om (RN) spacetime [see Eq. ({\ref{Eq14}) below]. 
In contrast to the Schwarzschild black hole, in these cases 
there is no simple one-to-one correspondence between the 
energy of the emitted quanta and the resulting change in black-hole surface area. 
Thus, the inclusion of chemical potentials may allow a deeper test of 
the applicability of Bohr's correspondence principle to the quantization of black holes. 
In this work we study the asymptotic resonances of a charged scalar field in the RN spacetime, 
and provide analytical formulae for the corresponding QNM spectrum. 
This is done by using a similarity between the QNMs of the charged scalar field 
and the known asymptotic spectrum of the natural field.

The dynamics of a charged scalar field in the RN spacetime is governed by the Klein-Gordon 
equation \cite{HodPir}

\begin{eqnarray}\label{Eq4}
\Delta {{d^2R_l} \over {dr^2}} +(2r-2M){{dR_l} \over {dr}}-l(l+1)R_l\nonumber \\
+{{r^4} \over \Delta}(\omega-{{eQ} \over r})^2R_l  = 0\  ,
\end{eqnarray}
where $\Delta \equiv (r-r_+)(r-r_-)$, and $r_{\pm} =M \pm (M^2-Q^2)^{1/2}$ are the 
black hole (event and inner) horizons. Here $e$ is the charge coupling constant [$e$ stands for $e/\hbar$, 
and has dimensions of (length)$^{-1}$].
The black hole QNMs correspond to
solutions of the wave equation with the physical boundary
conditions of outgoing waves at spatial infinity 
and ingoing waves crossing the event horizon \cite{Detwe}. Such boundary 
conditions single out a {\it discrete} set of resonances $\{\omega_n\}$. 
The solution to the wave equation may be expressed as 

\begin{eqnarray}\label{Eq5}
R_{l}& = &e^{i\omega r} (r-r_-)^{-1+i2M\omega+i\sigma_+} (r-r_+)^{-i\sigma_+}\nonumber \\
&&\Sigma_{n=0}^{\infty} d_n \Big({{r-r_+} \over {r-r_-}}\Big)^n\  ,
\end{eqnarray}
where $\sigma_{+} \equiv r^2_+(\omega-eQ/r_+)/(r_{+}-r_{-})$. 

The sequence of expansion coefficients $\{d_n:n=1,2,\ldots\}$ is determined by a 
recurrence relation of the form \cite{Leaver}

\begin{equation}\label{Eq6}
\alpha_n d_{n+1}+\beta_n d_n +\gamma_n d_{n-1}=0\  ,
\end{equation}
with initial conditions $d_0=1$ and $\alpha_0 d_1+\beta_0 d_0=0$. 
The quasinormal frequencies are determined by the requirement that the series in Eq. (\ref{Eq6}) is convergent, 
that is $\Sigma d_n$ exists and is finite \cite{Leaver}.

We find that the physical content of the recursion coefficients $\alpha_n,\beta_n$, and $\gamma_n$
becomes clear when they are expressed in terms of 
the black-hole physical parameters: the Bekenstein-Hawking temperature $T_{BH}=(r_{+}-r_{-})/A$, 
and the black-hole electric potential $\Phi=Q/r_{+}$, where $A=4\pi r_+^2$ is the black-hole surface area. 
The recursion coefficients obtain a surprisingly simple form in terms of these physical quantities, 

\begin{equation}\label{Eq7}
\alpha_n=(n+1)(n+1-2i\beta_{+}\hat\omega)\  ,
\end{equation}

\begin{eqnarray}\label{Eq8}
\beta_n& = &-2(n+{1 \over 2}-2i\beta_{+}\hat\omega)
(n+{1 \over 2}-2i\omega r_{+}+ieQ)\nonumber \\
&&-{1 \over 2} -l(l+1)\  ,
\end{eqnarray}
and

\begin{equation}\label{Eq9}
\gamma_n=[n-2i(2M\omega-eQ)](n-2i\beta_{+}\hat\omega)\  ,
\end{equation} 
where $\beta_{+} \equiv (4\pi T_{BH})^{-1}$ is the black-hole inverse temperature, and 
$\hat\omega \equiv \omega-e\Phi$.

We shall show that the quasinormal spectrum of a charged scalar field in the RN spacetime is closely related to 
the corresponding spectrum of a natural scalar field. 
The asymptotic spectrum of a natural scalar field in the RN spacetime is determined 
by the equation \cite{MotNei,Note2}

\begin{equation}\label{Eq10}
2e^{\mp 4\pi\beta_{+}\omega}+3e^{\mp 8\pi M\omega}=-1\ .
\end{equation}
Equation (\ref{Eq10}) suggests that the natural spectrum (for which $\hat\omega = \omega$) 
depends on the combinations $\beta_+ \omega$ 
and $2M\omega$ appearing in Eqs. (\ref{Eq7})-(\ref{Eq9}), but does not depend explicitly 
on $\omega r_+$. 
From Eqs. (\ref{Eq7})-(\ref{Eq9}) one learns that the analogy 
between the asymptotic spectrum of a natural scalar field and the 
corresponding spectrum of a charged field is obtained by applying the transformations 
$\beta_{+}\omega \to \beta_{+}\hat\omega$ and $2M\omega \to 2M\omega-eQ$ in Eq. (\ref{Eq10}). 
Using these transformations, one finds that the asymptotic quasinormal mode spectrum of 
a charged scalar field is given by 

\begin{equation}\label{Eq11}
2e^{\mp 4\pi\beta_{+}(\omega-e\Phi)}+3e^{\mp 4\pi(2M\omega-eQ)}=-1\  .
\end{equation}
For charged black holes that satisfy the condition $eQ \gtrsim r_+/r_-$ [this 
condition also reads $Q/M \gtrsim (\hbar/\alpha A)^{1/6}$, where $\alpha$ is 
the fine structure constant], one of the exponents (depending on the 
sign of $eQ$) in Eq. (\ref{Eq11}) 
is negligible as compared to the other, thus yielding two families of QNM resonances

\begin{equation}\label{Eq12}
\omega = \pm T_{BH}\ln2 +{{eQ} \over {r_+}} -i 2\pi T_{BH} (n+{1\over2})\  ,
\end{equation}
and 

\begin{equation}\label{Eq13}
\omega = \mp T^{s}_{BH}\ln3 +{{eQ} \over {r^s_+}} -i 2\pi T^{s}_{BH} (n+{1\over2})\  ,
\end{equation}
where $r^s_+=2M$ is the Schwarzschild radius, and 
the upper/lower signs correspond to positive/negative values of $eQ$, respectively \cite{Note3}.

The emission of a quantum of frequency $\omega$ and an electric charge $e$ results in a 
change $\Delta M=\hbar \omega_R$ in the black-hole mass, 
and a change $\Delta Q=e$ in its charge. 
Substituting the fundamental resonance, Eq. (\ref{Eq12}), into 
the first law of black-hole thermodynamics 

\begin{equation}\label{Eq14}
\Delta M={1 \over 4}T_{BH} \Delta A + \Phi \Delta Q\  ,
\end{equation}
one obtains the corresponding change in black-hole surface area

\begin{equation}\label{Eq15}
\Delta A =4\hbar \ln2\  .
\end{equation}
Remarkably, this fundamental change in black-hole surface area 
is in accord with the Bekenstein-Mukhanov general prediction, Eq. (\ref{Eq2}). 
In particular, it should be stressed that this area spacing turns out to be 
independent of the black-hole parameters, $M$ and $Q$, and also independent of 
the charged-field parameters, $e$ and $l$. The physical interpretation of the second branch, 
Eq. (\ref{Eq13}), is yet to be revealed.

In summary, motivated by novel results in the theory of black-hole quantization, 
we have studied analytically the QNM spectrum of a charged scalar field in the charged RN spacetime. 
It was shown that the asymptotic resonances can be expressed in terms of the black-hole physical parameters: 
its temperature $T_{BH}$, and its electric potential $\Phi$. 

The case of a {\it charged} field is interesting from a physical point of view, since it introduces 
a chemical potential into the system (in the form of the black-hole electric potential $\Phi$). 
This enabled us to test the applicability of Bohr's correspondence principle to the 
quantization of black holes in generalized situations, in 
which there is no one-to-one correspondence between the energy of the emitted quantum and the resulting 
change in black-hole surface area. 
We have shown that according to the Bohr correspondence principle, the emission of a charged quantum from a charged 
RN black hole induces a fundamental change in black-hole surface area, $\Delta A=4\hbar\ln2$. 
Remarkably, this area unit is universal in the sense that it is {\it independent} of the black-hole 
parameters, nor on the charged-field parameters.

\bigskip
\noindent
{\bf ACKNOWLEDGMENTS}
\bigskip

This research was supported by G.I.F. Foundation. I thank Uri Keshet for numerous discussions, as well as for 
a continuing stimulating collaboration.


\begin{thebibliography}{99}

\bibitem{Bekenmar} J. D. Bekenstein, 
Proceedings of the VIII Marcel Grossmann Meeting, T. Piran and R. Ruffini, eds. 
(World Scientific Singapore 1999). pp. 92-111.

\bibitem{Beken1} J. D. Bekenstein, Phys. Rev. D {\bf 7}, 2333 (1973).

\bibitem{Beken2} J. D. Bekenstein, Lett. Nuovo Cimento {\bf 11}, 467 (1974).

\bibitem{Ehren} See for example M. Born, Atomic Physics (Blackie,
  London, 1969), eighth edition.

\bibitem{Hod1} S. Hod, Class. Quant. Grav., {\bf 21}, L97 (2004).

\bibitem{Muk} V. Mukhanov, JETP Lett. {\bf 44}, 63 (1986).

\bibitem{BekMuk} J. D. Bekenstein and V. F. Mukhanov, Phys. Lett. B
  {\bf 360}, 7 (1995).

\bibitem{Beken3} J. D. Bekenstein in XVII Brazilian National Meeting
  on Particles and Fields, eds. A. J. da Silva et. al. (Brazilian
  Physical Society, Sao Paulo, 1996), J. D. Bekenstein in Proceedings of the VIII
  Marcel Grossmann Meeting on General Relativity, eds. T. Piran and
  R. Ruffini (World Scientific , Singapore, 1998).

\bibitem{Hod2} S. Hod, Phys. Rev. Lett. {\bf 81}, 4293 (1998).

\bibitem{Nollert1} For an excellent review and a detailed list of references see 
H. P. Nollert, Class. Quantum Grav. {\bf 16}, R159 (1999).

\bibitem{Press} W. H. Press, Astrophys. J. {\bf 170}, L105 (1971).

\bibitem{Cruz} V. de la Cruz, J. E. Chase and W. Israel,
  Phys. Rev. Lett. {\bf 24}, 423 (1970).

\bibitem{Vish} C.V. Vishveshwara, Nature {\bf 227}, 936 (1970).

\bibitem{Davis} M. Davis, R. Ruffini, W. H. Press and R. H. Price,
  Phys. Rev. Lett. {\bf 27}, 1466 (1971).

\bibitem{Leaver} E. W. Leaver, Proc. R. Soc. A {\bf 402}, 285 (1985).

\bibitem{Bach} A. Bachelot and A. Motet-Bachelot, Ann. Inst. H. Poincar\'e {\bf 59}, 3 (1993).

\bibitem{Nollert2} H. P. Nollert, Phys. Rev. D {\bf 47}, 5253 (1993).

\bibitem{Andersson} N. Andersson, Class. Quantum Grav. {\bf 10}, L61 (1993).

\bibitem{Note1} To understand the original argument, it is useful to recall that 
in the early development of quantum mechanics, Bohr 
suggested a correspondence between classical and quantum properties of the
Hydrogen atom, namely that ``transition frequencies at large quantum numbers should equal
classical oscillation frequencies''. The black hole is in many senses 
the ``Hydrogen atom'' of general relativity. It was therefore suggested \cite{Hod2} to use the 
{\it discrete} set of black-hole resonances in order to shed some light on the {\it quantum} 
properties of a black hole. However, there is one important difference between the Hydrogen atom and 
a black hole: while a (classical) atom emits radiation spontaneously according to the (classical) 
laws of electrodynamics, a {\it classical} black hole does not emit radiation. This crucial 
difference hints that one should look for the highly damped black-hole oscillations 
(let $\omega=\omega_R-i\omega_I$, then $\tau \equiv \omega_I^{-1}$ is the effective 
relaxation time for the black hole to return to a quiescent state after emitting radiation. Hence, the relaxation time $\tau \to 0$ as $\omega_I \to \infty$, implying no radiation emission, as should be the case for a classical black hole). Note also that the asymptotic real value of the Schwarzschild black-hole resonances is universal in the sense that it is independent of the harmonic index $l$, as expected if it is to describe a fundamental characteristic of the black hole itself.

\bibitem{Motl} L. Motl, Adv. Theor. Math. Phys. {\bf 6}, 1135 (2003).

\bibitem{Dreyer} O. Dreyer, Phys. Rev. Lett. {\bf 90}, 081301 (2003).

\bibitem{Ash} A. Ashtekar, J. C. Baez, A. Corichi and K. Krasnov, 
Phys. Rev. Lett. {\bf 80}, 904 (1998).

\bibitem{Rov} C. Rovelli and P. Upadhya, e-print gr-qc/9806079.

\bibitem{HodKesh} S. Hod and U. Keshet, Class. Quant. Grav. {\bf 22}, L71 (2005).

\bibitem{HodPir} S. Hod and T. Piran, Phys. Rev. D {\bf 58}, 024017 (1998).

\bibitem{Detwe} S. L. Detweiler, in Sources of Gravitational 
Radiation, edited by L. Smarr (Cambridge University Press, 
Cambridge, England, 1979).

\bibitem{MotNei} L. Motl and A. Neitzke, Adv. Theor. Math. Phys. {\bf 7}, 307 (2003).

\bibitem{Note2} The final result of \cite{MotNei} for the asymptotic QNM spectrum of 
{\it natural} fields in the RN spacetime, can be written as Eq. (\ref{Eq10}) by simply multiplying both sides of 
equation $(66)$ in \cite{MotNei} by $e^{-\beta_{+} \omega}$.

\bibitem{Note3} Note that these solutions have the desired symmetry, namely that if 
$\omega(eQ)$ is a QNM frequency, then $-\omega^*(-eQ)$ is also a solution.

\end{thebibliography}
\end{document}